\documentclass[]{eptcs}
\usepackage{courier}
\usepackage[latin9]{inputenc}
\usepackage{listings}
\usepackage{rotating}
\usepackage{url}
\usepackage{graphicx}

\makeatletter

\providecommand{\tabularnewline}{\\}

\providecommand{\event}{TTC 2013} 

\def\titlerunning{Petri-Nets to Statecharts}
\def\authorrunning{P. {Van Gorp} and L.M. Rose}

\usepackage{url}\usepackage{color}\usepackage{listings}

\title{The Petri-Nets to Statecharts Transformation Case}

\author{Pieter Van Gorp \institute{School of Industrial Engineering, Eindhoven University of Technology, The Netherlands.} \email{p.m.e.v.gorp@tue.nl} \and Louis M. Rose \institute{Department of Computer Science, University of York, UK.} \email{louis.rose@york.ac.uk}}

\makeatother

\begin{document}
\maketitle

This paper describes a case study for the sixth Transformation Tool
Contest. The case is based on a mapping from Petri-Nets to statecharts
(i.e., from flat process models to hierarchical ones). The case description
separates a simple mapping phase from a phase that involves the step
by step destruction Petri-Net elements and the corresponding construction
of a hierarchy of statechart elements. Although the focus of this
case study is on the comparison of the runtime performance of solutions,
we also include correctness tests as well as bonus criteria for evaluating
transformation language and tool features.

\section{Introduction}

Although transformations are already developed for decades in various
communities (such as the compiler community~\cite{Bacon1994compilertrans},
the program transformation community~\cite{Feather87progtranssurv}
and the business process management (BPM) community~\cite{Lohmann09BPMtransSurv}),
it is relatively new to study the strengths and weaknesses of transformation
approaches across community boundaries. The aim of the Transformation
Tool Contest (TTC) is to compare the expressiveness, the usability
and the performance of graph and model transformation tools along
a number of selected case studies. A deeper understanding of the relative
merits of different tool features will help to further improve graph
and model transformation tools and to indicate open problems~\cite{TTC11Proceedings}.
This paper proposes a case for the sixth edition of TTC (i.e., to
TTC'13).

\subsection{Context of the Case}

In BPM research and practice, transformations are usually programmed
using general purpose programming languages. Mapping rules in BPM
literature tend to be formalized using mathematical set constructs
whereas they tend to be documented using informal visual rules and
implemented in Java or another general purpose programming language.
In the graph and model transformation communities, special purpose
languages and tools are being developed to support the direct execution
of such mapping rules. Finally, various comparative studies have been
contributed to the emerging field of transformation engineering (cfr.,~\cite{Taentzer2005_ModelTransformationsbyGraphTransformationsAComparativeStudy,Varro2008_TransformationofUMLModelstoCSPACaseStudyforGraphTransformationTools,Rensink2010_Graphtransformationtoolcontest2008,Gronmo2009_ComparisonofThreeModelTransformationLanguages,Rose2009_AnAnalysisofApproachestoModelMigration,rose10comparison,vAmstel2011ATLQVTperfAnal,Tolosa2011TSMatlMeasurement})
but too little transformation problems have been considered that are
considered challenging by the BPM community. This paper proposes the
so-called PN2SC case, related to the transformation of Petri-Nets
to statecharts. The case relates to an active research line at Eindhoven
University of Technology. Interestingly, much of the associated research
efforts have been spent on an optimization algorithm which turns out
to be irrelevant when using a rule-based transformation approach instead
of an imperative programming approach~\cite{VanGorp2010MoDELS}.
Therefore, the case can be considered a nice showcase for demonstrating
to the BPM community the potential impact of results from the transformation
community.

\subsection{Relevance for TTC'13}

Besides providing 
input to BPM industry and its research community, this study also
covers a previously unstudied type of transformation. More specifically,
the transformation problem under study is a process model translation
that raises the abstraction level. Regardless of concrete language
and metamodel details, this type of problem has not yet been considered
for the evaluation of transformation approaches. 

In the following, we briefly survey existing comparative studies related
to process models. Varró et al. studied various approaches to mapping
conceptual process model in UML to more technical CSP models~ \cite{Varro2008_TransformationofUMLModelstoCSPACaseStudyforGraphTransformationTools}.
Also, in a special issue edited by Rensink et al., various experts
present their solutions to a case study concerning the mapping of
conceptual process models in BPMN to more technical BPEL models~\cite{Rensink2010_Graphtransformationtoolcontest2008}.
Grønmo et al. discuss various approaches to transforming conceptual
UML models into strictly structured counter-parts~\cite{Gronmo2009_ComparisonofThreeModelTransformationLanguages}.
Again, the target models are more low-level than the input models.
Rose et al. discuss different approaches to migrating Petri-Net models~\cite{rose10comparison}.
In that case, input models are at the same level of abstraction as
output models. 

Van Gorp et al. have already considered the case proposed by this
paper for comparing a graph transformation approach to the Java-based
programming approach that is mainstream in the BPM domain. The authors
make the interesting observation that the rule-based approach required
less specification effort yet delivered superior performance. By submitting
this case to TTC'13, we aim at analyzing the generalizability of that
observation.

\section{The Transformation}

\label{sec:TheTransfo}
\begin{figure}[tb]
\centerline{ %
\begin{tabular}{c||c}
\includegraphics[width=0.45\linewidth]{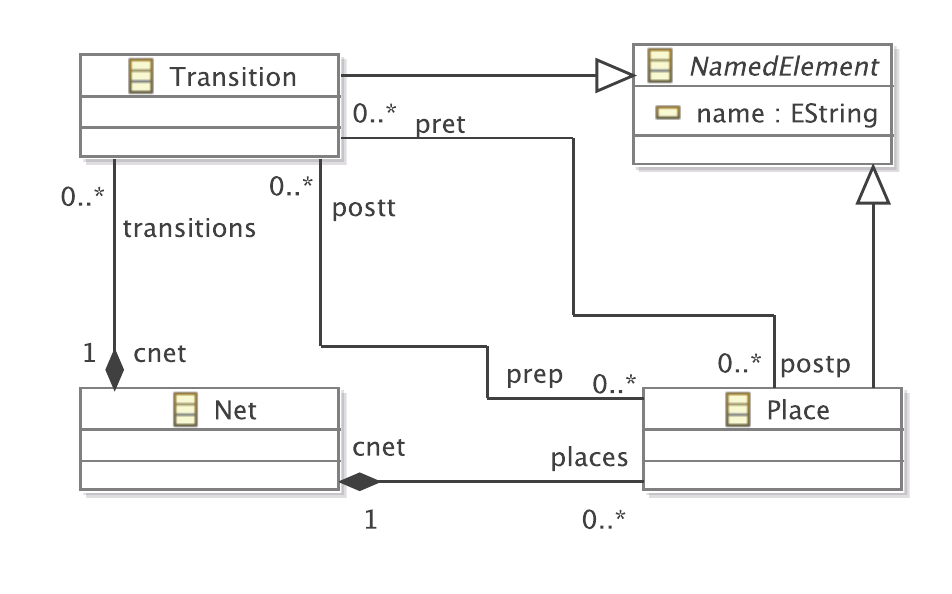}  & \includegraphics[width=0.5\linewidth]{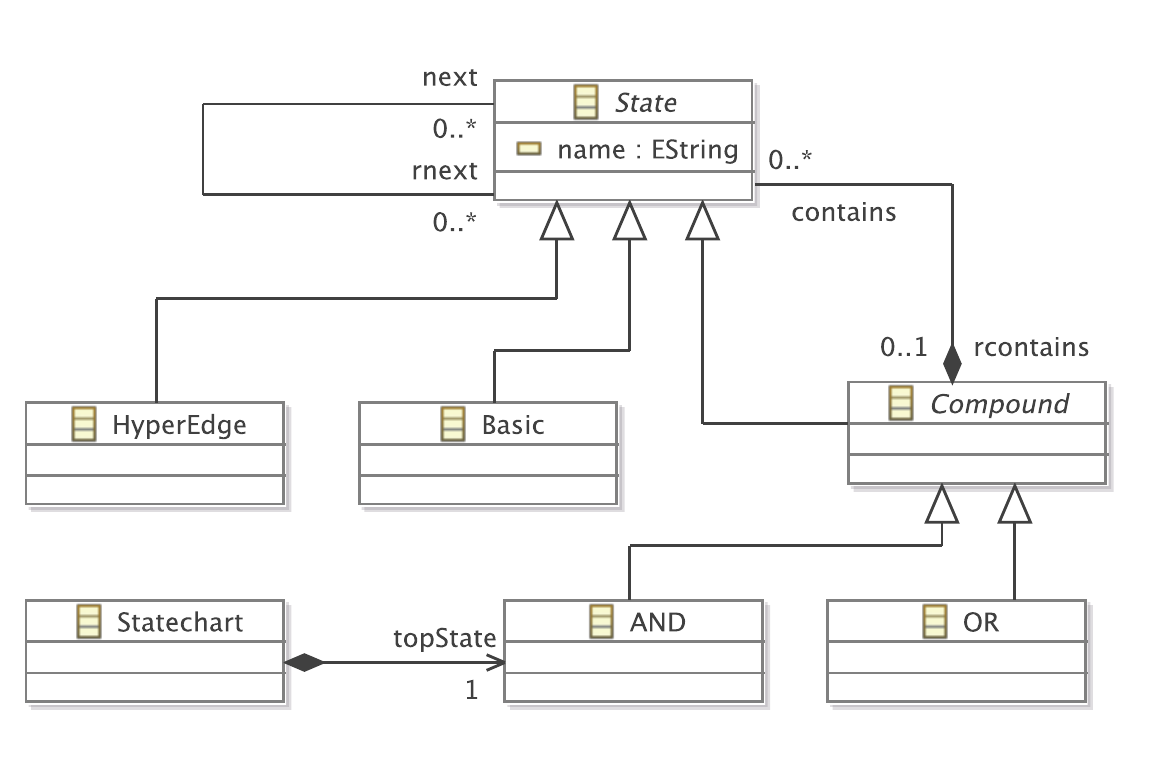}\tabularnewline
(a) Petri-Net metamodel  & (b) Statechart metamodel \tabularnewline
\end{tabular}} \caption{\label{fig:mms}The metamodels used in the PN2SC transformation.}
\end{figure}

This section details the Petri-Net to statechart
transformation algorithm, originally described by Eshuis~\cite{eshuis05pn2scTechreport}.
We summarise the key details of the transformation below, and strongly
encourage solution developers to refer to \cite{eshuis05pn2scTechreport} for
a more comprehensive discussion of the transformation and the rationale behind
the transformation rules. 
The transformation described below is \emph{input-destructive} (elements
of the Petri-Net model are removed as the transformation proceeds),
and uses the metamodels shown in figure~\ref{fig:mms}. Variants
of the rules can be designed such that the input model is not destroyed.
However, in order to make all solutions comparable at the level of
languages and tools, discourage unnecessary variations at the solution
design level. In general, we encourage solution submitters to stick
as closely as possible to the conceptual requirements, as described
in the following.

\subsection{Preconditions}

The following assumptions can be made regarding the context in which
the PN2SC transformation is started: 
\begin{itemize}
\item there is only one instance of \texttt{Net},
\item there are no \texttt{Statechart} nor \texttt{State} instances.
\end{itemize}

\subsection{Initialization}

The first step in the PN2SC transformation involves creating an initial
structure for the statechart model. In particular, the following statechart
model elements are created:
\begin{itemize}
\item For every \texttt{Place} $p$ in the Petri-Net: \subitem an instance
of \texttt{Basic}, $b$ (with $b.name=p.name$), \subitem an instance
of \texttt{OR}, $o$ such that $o.contains=\{b\}$,
\item For every \texttt{Transition} $t$ in the Petri-Net: \subitem an
instance of \texttt{HyperEdge} $e$ (with $e.name=t.name$),
\item All \texttt{pret/postp} and \texttt{postt/prep }arcs should be mapped
to \texttt{next/rnext} links between the \texttt{States} equivalent
with the input \texttt{NamedElements} that are connected by these
arcs.
\end{itemize}

\paragraph{Equivalence.}

Initialization should also provide a mechanism for identifying the
\texttt{OR} node created for a particular \texttt{Place} and the \texttt{HyperEdge}
created for a particular \texttt{Transition}. The precise mechanism
can vary over implementations. One approach is to use a name-based
identification (e.g., assign all \texttt{Places} a uniquely identifying
name and copy each \texttt{Place's} name to its \texttt{OR} node during
initialization.) Another approach is to create traceability links
between corresponding source and target elements. In the remainder
of this section we assume that the initialization of the transformation
will construct an injective function $equiv:Place\to OR$.

\subsection{Reduction rules}

Following initialization, the transformation continues by applying
one of two types of reduction rules: AND and OR. 

\begin{figure}[tb]
\centerline{ %
\begin{tabular}{c||c}
\includegraphics[width=0.45\linewidth]{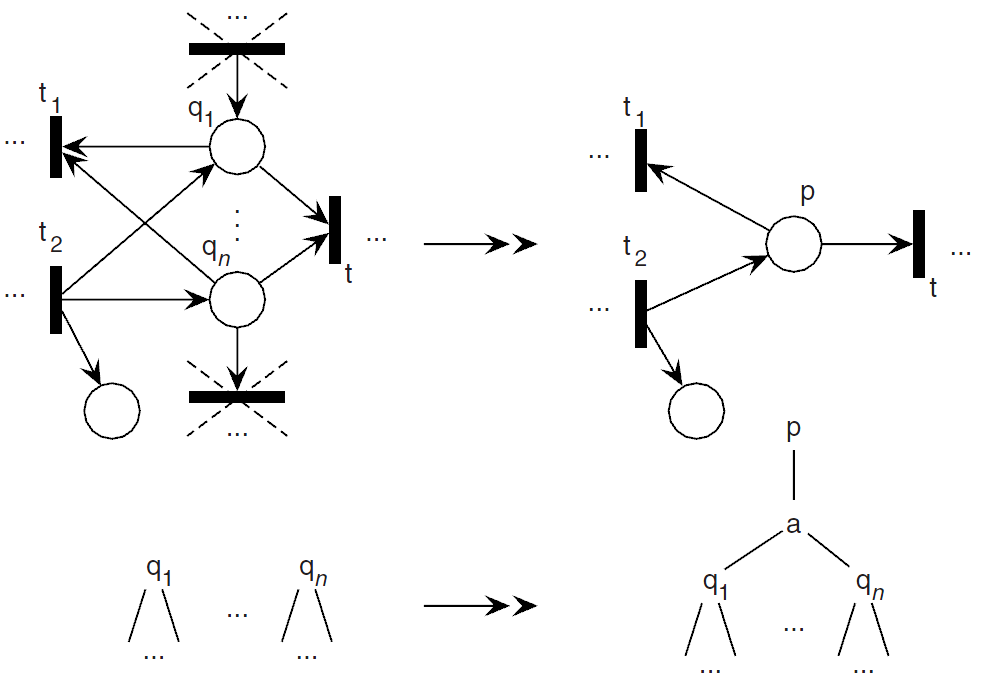}  & \includegraphics[width=0.5\linewidth]{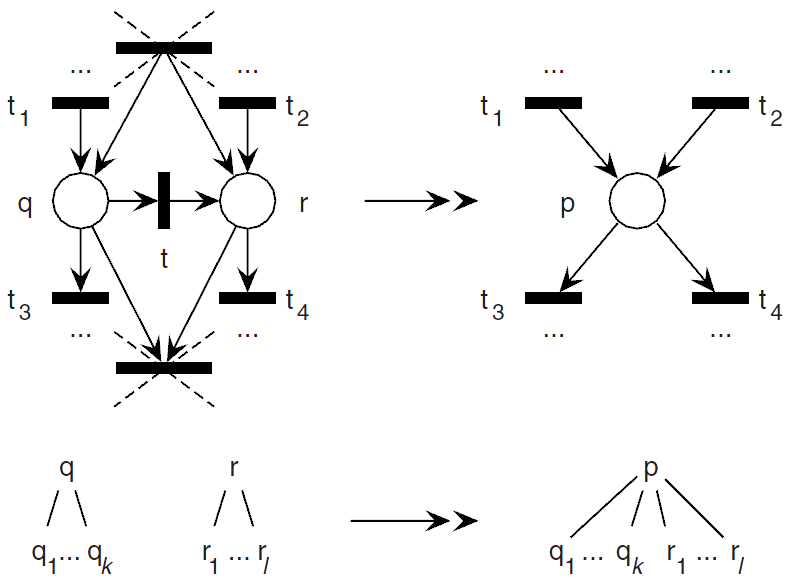}\tabularnewline
(a) rule for creating \emph{AND nodes}  & (b) rule for merging \emph{OR nodes}\tabularnewline
\end{tabular}} \caption{\label{fig:Visual-documentation-for-rules}Visual documentation for
mapping rules.}
\end{figure}

\subsubsection{AND rules}

The first type of reduction rule, AND (informally documented by Figure~\ref{fig:Visual-documentation-for-rules}.a),
constructs an \texttt{AND} state for a set of \texttt{Places} that
are connected to the same incoming and outgoing \texttt{Transitions}.

\paragraph{Pre-conditions.}

The AND rule can be applied to a \texttt{Transition}, $t$, iff $\left|t.prep\right|>1$
and every \texttt{Place} in $t.prep$ is connected to the same set
of outgoing transitions and the same set of incoming transitions.
Alternatively, the AND rule can be applied to a \texttt{Transition},
$t$ iff $\left|t.postp\right|>1$ and every \texttt{Place} in $t.postp$
is connected to the same set of outgoing transitions and the same
set of incoming transitions. Given these two alternative situations,
we refer to two AND rules. However, in some languages one can implement
these variants in one rule.

Figure~\ref{fig:Visual-documentation-for-rules}.a illustrates when
the AND rule for the situation $\left|t.prep\right|>1$ would be applicable:
$q_{1}$ to $q_{n}$ all have $\{t,t_{1},...\}$ as the set of outgoing transition,
and they share $\{t_{2},...\}$ as the set of incoming transitions.
$q_{n}$ should not have an outgoing arc to a transition for which
there is no corresponding arc between $q_{1}$ and that transition.
Conversely, $q_{1}$ should not have an incoming arc from a transition
for which there is no corresponding arc between that transition and
$q_{n}$. To illustrate the situation when the AND rule for the situation
$\left|t.postp\right|>1$ would be applicable, one can simply reverse
all arcs in Figure~\ref{fig:Visual-documentation-for-rules}.a.

\paragraph{Effect on statechart.}

Applying the AND rule results in the creation of a new \texttt{AND}
state ($a$) and a new \texttt{OR} state ($p$) such that $p.contains={a}$
and $a.contains$ is the set of \texttt{OR} states $\{q\in t.prep:equiv(q)\}$;
or $\{q\in t.postp:equiv(q)\}$ if the rule has been applied to the
transition's postset, $t.postp$.

\paragraph{Effect on Petri-Net.}

Applying the AND rule removes from the Petri-Net all but one of the
\texttt{Places} in the set $t.prep$ ($t.postp$).

\paragraph{Example Implementations.}

An EOL implementation of the AND rule can be found online\footnote{See \url{https://github.com/louismrose/ttc_pn2sc/blob/master/example_solution/reduction/and.eol}}.
That EOL solution applies to the cases $\left|t.prep\right|>1$ and $\left|t.postp\right|>1$ without requiring code duplication.  A GrGen.NET implementation based on two separate variants of the AND rule can also be found online\footnote{See \url{https://github.com/pvgorp/TTC13-PN2SC-GrGen/blob/master/ttc13-version/PNtoHSC.grg} from line 109 to line 149.}.

\subsubsection{OR rules}

The second type of reduction rule, OR (informally documented by Figure~\ref{fig:Visual-documentation-for-rules}.b),
constructs an \texttt{OR} state for a \texttt{Transition} that has
a single preceding \texttt{Place} and single succeeding \texttt{Place}.

\paragraph{Pre-conditions.}

The OR rule can be applied to a \texttt{Transition}, $t$, iff $(\left|t.prep\right|=1)\land(\left|t.postp\right|=1)$
and there is no transition, $t'$, such that $(q\in t'.prep)\land(r\in t'.prep)$
or $(q\in t'.postp)\land(r\in t'.postp)$ where $q$ is the single
place contained in $t.prep$ and $r$ is the single place contained
in $t.postp$.

\paragraph{Effect on statechart.}

Applying the OR rule results in the creation of a new \texttt{OR}
state ($p$) such that $p.contains$ is the set of \texttt{OR} states
$equiv(q).contains\cup equiv(r).contains$.

\paragraph{Effect on Petri-Net.}

Applying the OR rule removes from the Petri-Net the \texttt{Transition}
$t$ and the \texttt{Places} $q$ and $r$; and adds a new \texttt{Place}
$p$ such that $p.pret=(q.pret\cup r.pret)$ and $p.postt=(q.postt\cup r.postt)$.

\paragraph{Example Implementations.}

An EOL implementation of the OR rule can be found online\footnote{See \url{https://github.com/louismrose/ttc_pn2sc/blob/master/example_solution/reduction/or.eol}}.
This implementation re-uses $q$ to form $p$,
rather than instantiate a new \texttt{Place} or \texttt{OR} state).  A GrGen.NET implementation can also be found online\footnote{See \url{https://github.com/pvgorp/TTC13-PN2SC-GrGen/blob/master/ttc13-version/PNtoHSC.grg} from line 154 to line 189.}.

\subsection{Termination}

The output model should contain a single instance of \texttt{Statechart}
$sc$. The transformation should apply the reduction rules as long
as possible to the input elements. For input Petri-Nets in the class
of nets defined by~\cite{eshuis09fm}, the reduction process should
terminate with exactly one \texttt{Place} left. The equivalent \texttt{AND}
state $st$ should not have any parent \texttt{Compound} state. Moverover,
$sc.topState=st$ should hold for $sc$. For input nets outside that
class, the reduction rules typically cannot produce a unique top-level
\texttt{AND} state so it is impossible to produce a model that conforms
strictly to the output model. However, in that case the transformation
should terminate in a state that can only be reached after sequentially
applying the reduction rules as long as possible. It should be possible
to inspect the resulting state of the input Petri-Net for further
analysis purposes.

\section{Example Transformation Execution Trace}

\begin{figure}

\begin{centering}
\includegraphics[height=0.8\paperheight]{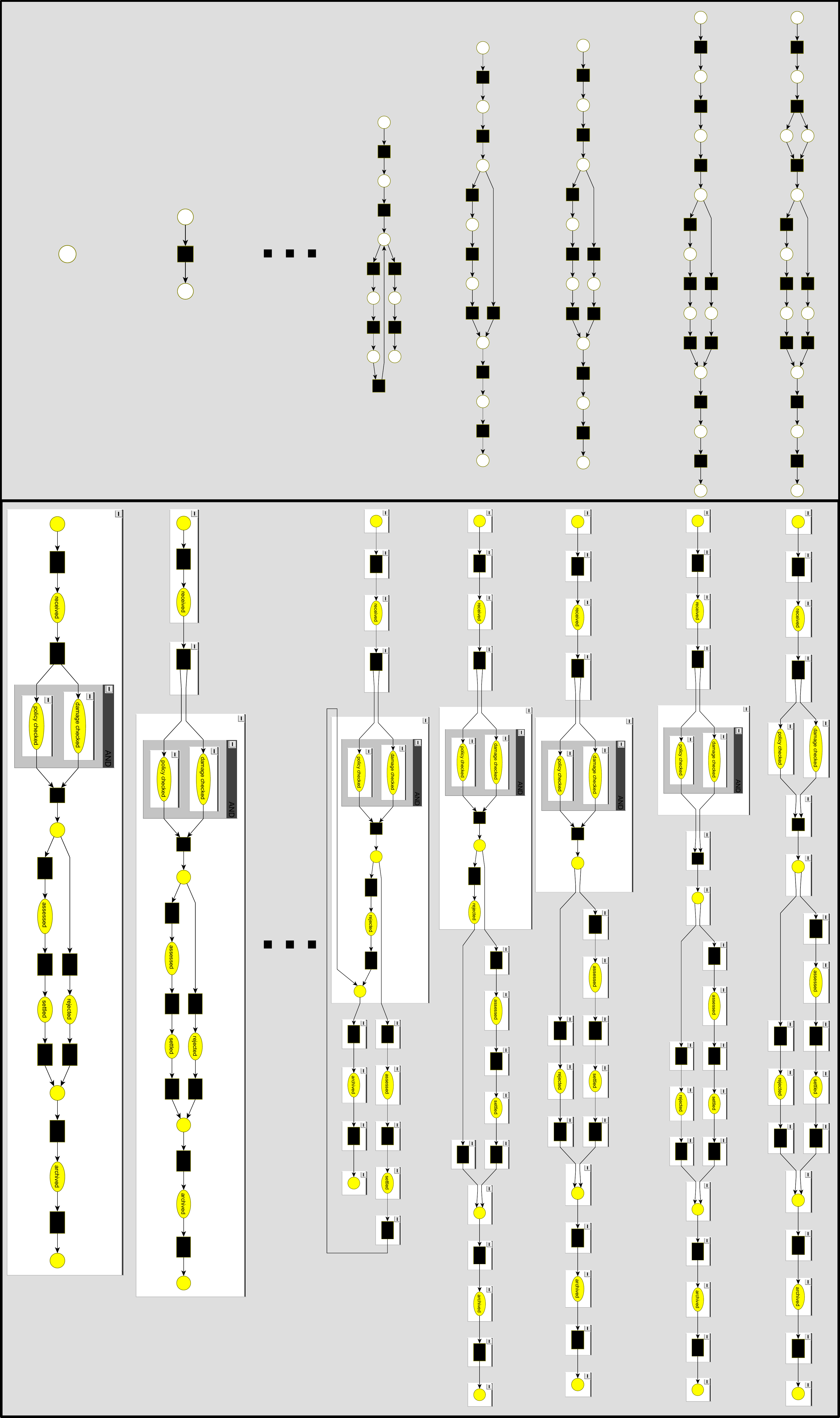}
\par\end{centering}

\caption{\label{fig:Visualization-ANDandOR}Visualization of an execution trace
of the AND and OR reduction rules.}

\end{figure}

This section demonstrates the intended behavior of the reduction rules
from the previous section. Figure~\ref{fig:Visualization-ANDandOR}
visualizes an execution trace for an input model with 11 places and
10 transitions. The input Petri-Net model is shown on the left of
the figure while its corresponding statechart is shown at the right
of the figure. Places are represented as white circles and transitions
as black bars. Basic states are represented as yellow ovals while
hyperedges are represented as black bars too. OR states are represented
as white boxes while AND states are represented as gray boxes. The
top of the figure shows an execution snapshot after initialization.
Each \texttt{Place} and \texttt{Transition} element has respectively
its corresponding \texttt{Basic} state and \texttt{HyperEdge}. 

The figure shows subsequent applications of reduction rules.
The second to fifth snapshots are the result of applying in sequence
once the AND rule and three times the OR rule. From the figure it
cannot be determined which variant of the AND rule has been applied
in the first step since they would both lead to the second snapshot
in Figure~\ref{fig:Visualization-ANDandOR}. The step from the second to the third snapshot (i.e., the first application of the OR rule) looks complex at the statechart side (i.e., at the right side of the figure) since there are multiple arcs arriving at the transition that is reduced.  However, it is straightfoward at the Petri-Net side (i.e., at the left side) and that is the side at which the reduction rules are matching against input elements.  The step from the fourth to the fifth snapshot does require some explanation even at the Petri-Net side: the fifth Petri-Net snapshot contains a loop, which may seem remarkable since there are no loops in the input Petri-Net.  Still, the snapshot is the result of correctly applying the OR rule to those two places in the Petri-Net that contain respectively two outgoing arcs and two incoming arcs.  Those two places are merged into one place, which has as set of pre- and post-arcs the respective unions of the sets of pre- and post-arcs of those two places.

The bottom of the figure
shows a final application of the OR rule. Since the final snapshot
contains exactly one input place, the output statechart can be considered
valid. The details of embedding the final structure in a StateChart
container element are straightforward and not shown in Figure~\ref{fig:Visualization-ANDandOR}.

\section{Testsuite and Additional Artifacts}

In this section, we present some testcases that should be used both
for some basic correctness verification as well as for performance
testing. The complete testsuite is available in an online repository%
\footnote{See \url{https://github.com/louismrose/ttc_pn2sc/}%
}. 
Moreover, we provide online access to additional test materials,
all three existing solutions to the case (a Java solution, a GrGen.NET
solution and an Epsilon solution%
\footnote{See \url{http://is.ieis.tue.nl/staff/pvgorp/share/?page=LookupImage&bNameSearch=pn2sc+benchmark}.%
}) and GMF-based editors with automatic layouting features. We have
also participated in online discussions related to these materials\footnote{See \url{http://goo.gl/EjyCwo}}.  In the context of these discussions, we have 

\subsection{Testcase 1: Artificial Example}

This testcase is based on the running example from a Formal Methods
2009 conference paper by Eshuis~\cite{eshuis09fm}. Some intermediate
snapshots of a valid application sequence of the reduction rules are
described in that paper but the distinction between input and output
model is not as clear as in Figure~\ref{fig:Visualization-ANDandOR}. 

\begin{figure}
\begin{centering}
\includegraphics[width=0.6\linewidth]{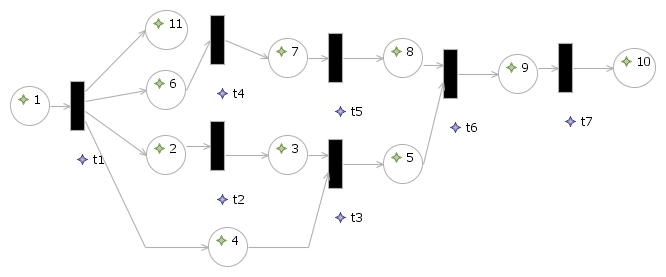}
\par\end{centering}

\caption{\label{fig:Testcase-1-pn}Testcase 1 input, diagram exported from
a GMF-based Petri-Net editor.}

\end{figure}

\begin{figure}
\begin{centering}
\includegraphics[width=1\linewidth]{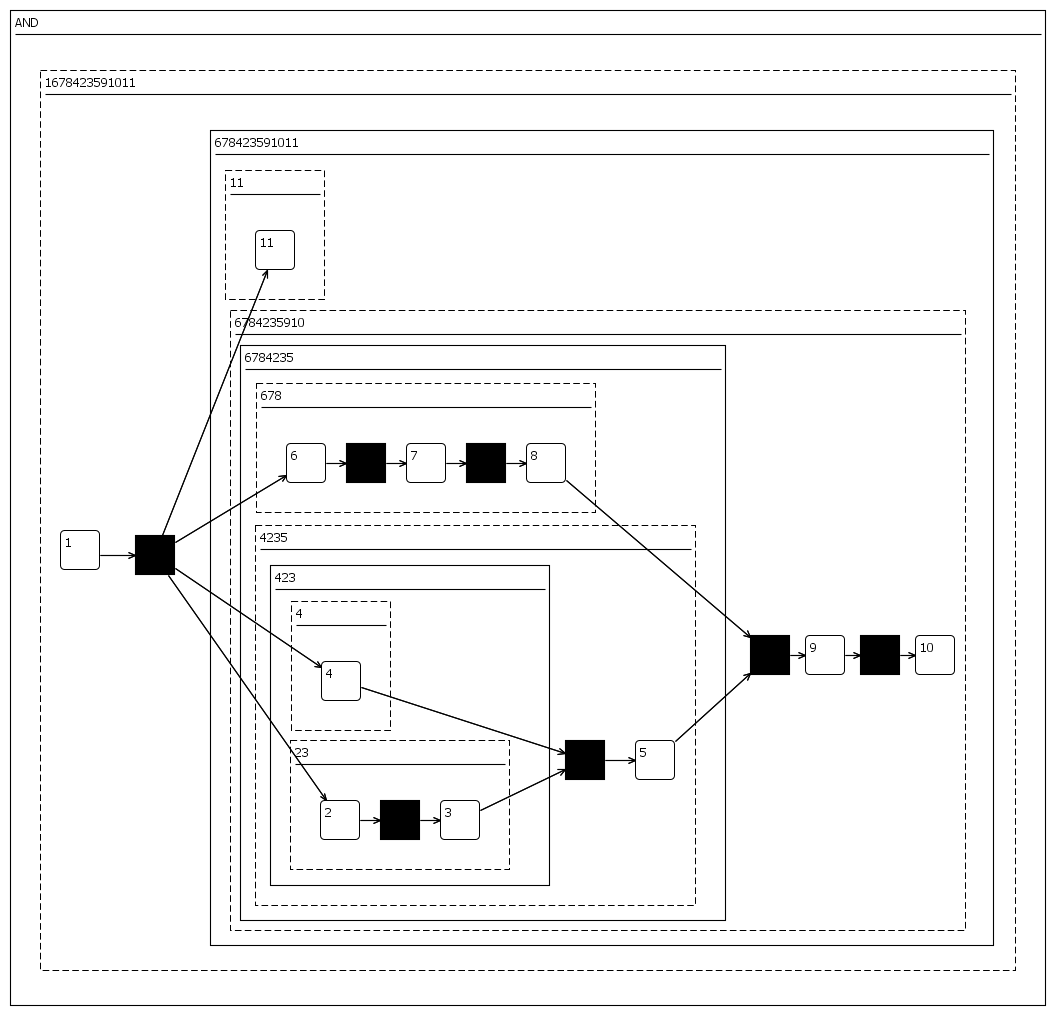}
\par\end{centering}

\caption{\label{fig:Testcase-1-out}Testcase 1 output, diagram exported from
a GMF-based Statechart editor.}
\end{figure}

Figure~\ref{fig:Testcase-1-pn} shows the input Petri-Net while Figure~\ref{fig:Testcase-1-out}
shows the expected output. The names of the AND and OR states are
arbitrary. Both figures are based on GMF-based editors so the details
of the concrete Petri-Net and statechart syntax can be adapted. In
the following, AND states are represented as rectangles with solid
lines while OR states are represented as rectangles with dashed lines.

\subsection{Testcase 2: Insurance Claim Example}

This testcase is based on an insurance claim example from a paper
at the ER 2012 conference~\cite{EshuisVanGorp2012ER}. For that paper,
PN2SC is supplemented with pre-processing rules for activity diagrams
with object flows. Also, there are post-processing rules for removing
hyper-edges and for introducing initial, decision and final nodes
in the statecharts. These extensions are not part of the proposed
TTC'13 case but we mention them here for clarifying the bigger picture.

\begin{figure}
\begin{centering}
\includegraphics[width=1\linewidth]{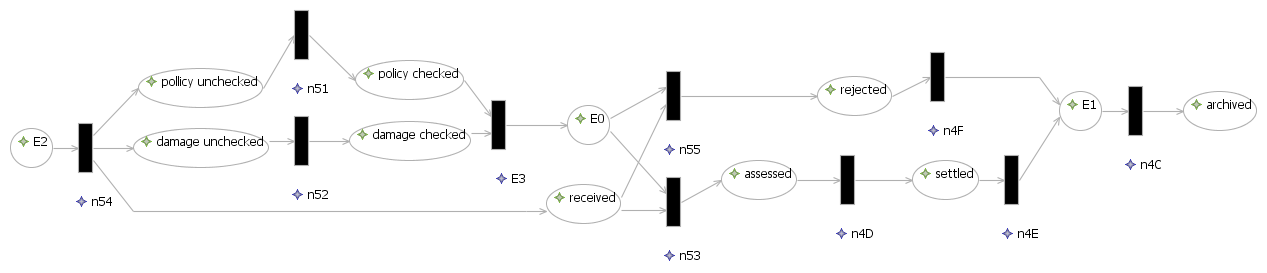}
\par\end{centering}

\caption{\label{fig:Testcase-2-pn}Testcase 2 input, diagram exported from
a GMF-based Petri-Net editor.}
\end{figure}

\begin{figure}
\begin{centering}
\includegraphics[width=1\linewidth]{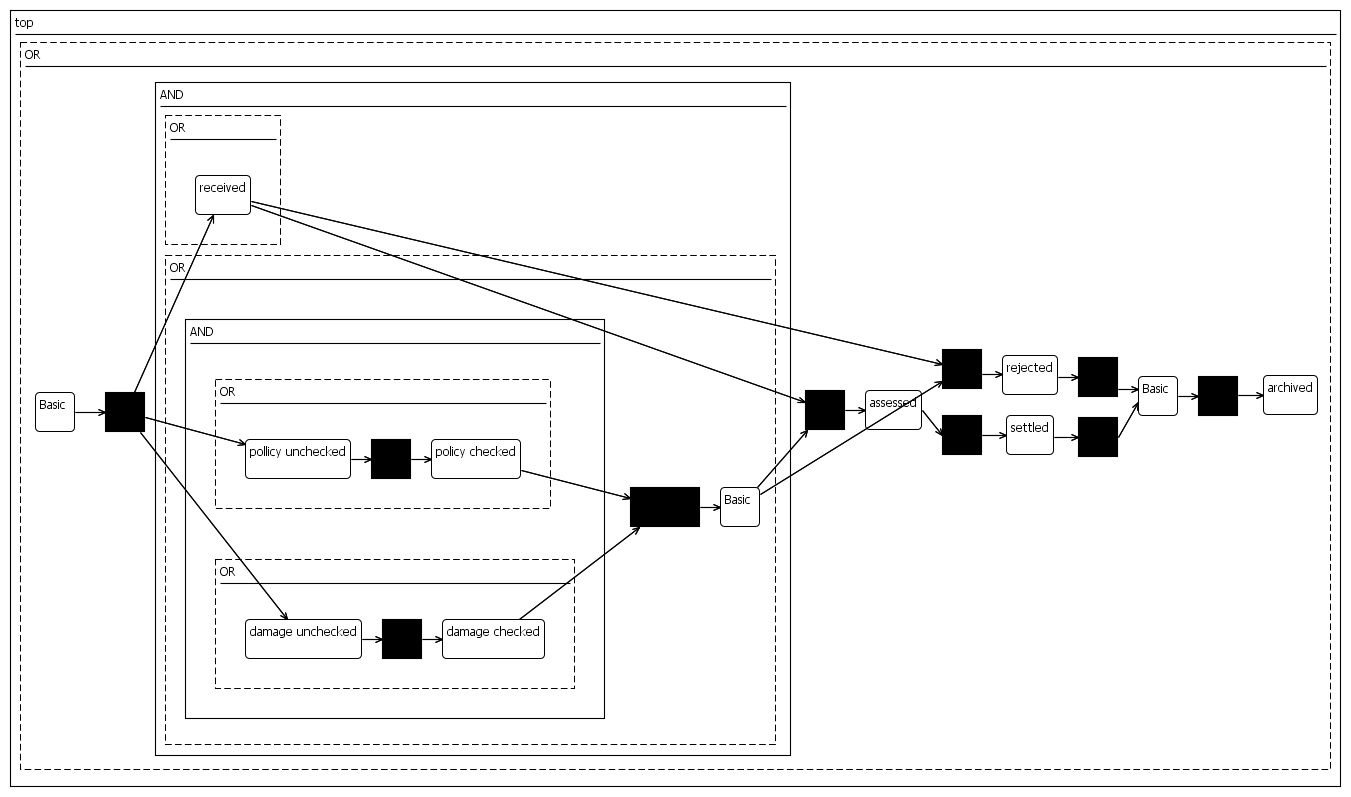}
\par\end{centering}

\caption{\label{fig:Testcase-2-out}Testcase 2 output, diagram exported from
a GMF-based Statechart editor.}
\end{figure}

Figure~\ref{fig:Testcase-2-pn} shows the input Petri-Net for Testcase
2 while Figure~\ref{fig:Testcase-2-out} shows the expected output.
Similar to Testcase 1, the input Petri-Net should be completely
reduced to one place upon transformation termination.

\subsection{Testcase 3: Patient Logistics Example}

This testcase is inspired by the way patients can be routed throughout
a hospital in the context of dermatology care. For this testcase,
the transformation should terminate in the aforementioned failure
state since the input model is not part of the class of Petri-Nets
for which the reduction rules have been designed.

\begin{figure}
\begin{centering}
\includegraphics[width=1\linewidth]{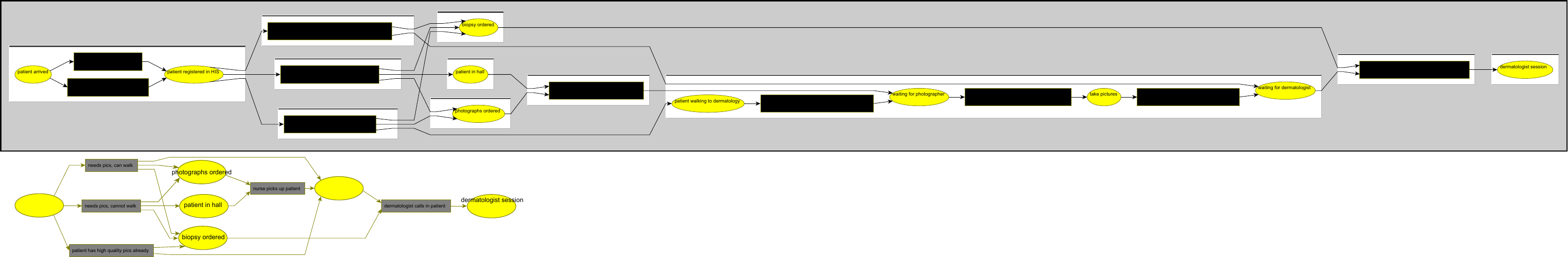}
\par\end{centering}

\caption{\label{fig:Testcase-3-inoutFail}Testcase 3 output, diagram exported
from the GrGen.NET graph visualizer.}
\end{figure}

Figure~\ref{fig:Testcase-3-inoutFail} shows the input Petri-Net for Testcase
3 as well as a possible visualization of the failed execution. The top of the figure shows the statechart
model that was generated so far. Again, OR states are represented
as white boxes. There are no AND states in the model, but a gray box
is shown nonethless to make the white boxes better visible in print.
The bottom of the figure shows the Petri-Net upon transformation failure.
The two places (ovals) without labels have been produced by successive
applications of the OR reduction rule. The corresponding OR states
can be identified at the left and in the middle of the statechart.

\subsection{Performance Testcases}

For performance testing, we enclose an EMF representations of the
key models that were used to compare the Java and GrGen.NET based
solutions of the aforementioned paper by Van Gorp et al.~\cite{VanGorp2010MoDELS}.
The models are based on a large Petri-Net that can be transformed
successfully by the reduction rules. That Petri-Net has been cloned and merged
multiple times to create models of up to 300 thousand elements. All
results from~\cite{VanGorp2010MoDELS} can be reproduced via an online
virtual machine in SHARE~\cite{vangorp10PNtoHSCiiii}. In general,
the performance of both solutions scales linearly for models of normal
size (requiring always less than a second.) For models with thousands
of elements, the GrGen.NET solution exhibits $\bigcirc(x^{2})$
time complexity.

\begin{figure}
\begin{centering}
\begin{tabular}{c}
  \includegraphics[width=0.62\linewidth]{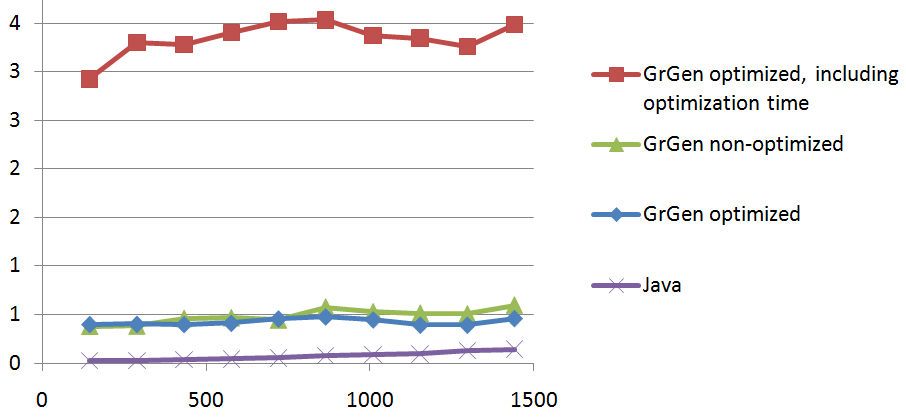} \tabularnewline
  \multicolumn{1}{c}{(a)}
\end{tabular}
\begin{tabular}{c|c}
 \includegraphics[width=0.35\linewidth]{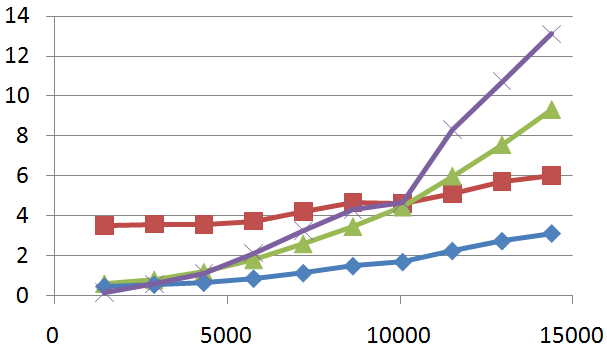} & \includegraphics[width=0.35\linewidth]{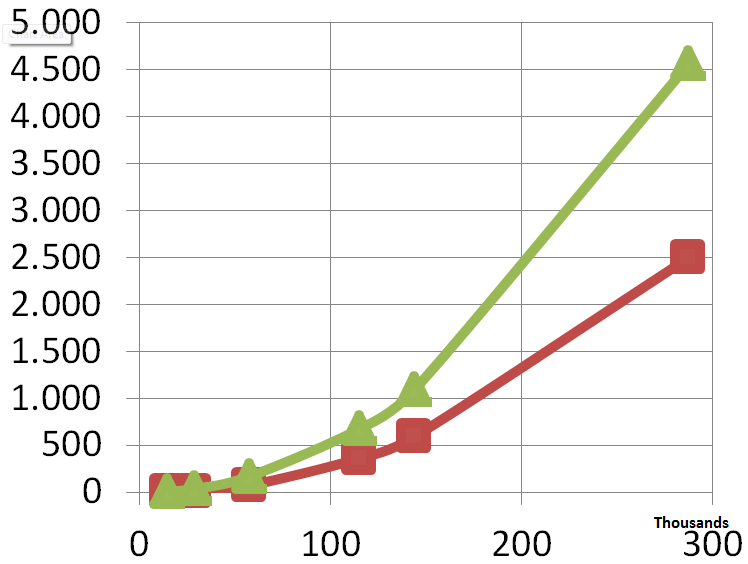}\tabularnewline
\multicolumn{1}{c}{(b)} & (c)\tabularnewline
\end{tabular}
\par\end{centering}

\centering{}\caption{\label{fig:perf} Performance (execution time seconds) of Java and
GrGen.NET solutions.}
\end{figure}

Figure~\ref{fig:perf} shows the performance results from a benchmark~\cite{VanGorp2010MoDELS}.
The figure shows that for GrGen.NET, two curves are shown. Both curves
are based on just one transformation specification. Different results
were obtained by changing small engine settings: for one version,
the rules are applied directly while for the other version the engine
is instructed to first analyze the input model and generate optimized
execution code/parameters. The analysis obviously takes time too.
It turned out that for models consisting of less than 10.000 elements,
the analysis generated too much overhead. The Java solution cannot
process more than about 15.000 elements, due to limitations of the
address space of the 32 bits Java virtual machine that we have used.
For inputs models with more than 10.000 elements, the required processing
time increases significantly too. For models of some hundreds of thousands
of elements, the engine-provided GrGen optimizations reach a speedup
factor of almost two.

Note that our performance tuning and testing of the Epsilon solution described earlier in this paper is not yet complete. Our early results indicate that the Epsilon solution is one or two orders of magnitude slower than the Java solution, which we attribute to a lack of performance tuning, and the interpretative nature of EMF and the Epsilon Object Language.

We encourage case submitters to demonstrate various strategies to performance
tuning but point out that for the GrGen.NET solution, the transformation
definition (i.e., the rule specification) was not even optimized for
performance. In general, especially solutions that deliver good performance
without negatively affecting the specification style are considered
of high quality.

Measurements taken for the purposes of comparing performance
with other solutions must include two intervals: time taken to load / import
models and time take to execute the transformation on a model once it has been
loaded and imported. To ensure an equitable comparison, measurements must be taken
 on a SHARE virtual machine image.

\subsection{Testcases 4-11: Advanced Control Flows}
During the solution discussion phase, we observed the need to share additional testcases.  In particular, the testsuite had to be extended with inputs involving more advanced control flow structures.  Three weeks before the solution submission deadline, we have shared 8 additional testcases\footnote{See \url{http://goo.gl/U1ksGQ}}.  
The additional testcases involve fabricated Petri-Nets that contain advanced Petri-Net synchronization and loop structures.  Not all additional petri-nets can be reduced successfully.  In summary, input models 1,2, 7, 8 and 11 should be successfully reduced to statecharts while nets 3, 5, 6, 9 and 10 cannot (and should not) be reduced completely.

\section{Evaluation Criteria}

\subsection{Basic Criteria}

\subsubsection{Transformation Correctness}

Solutions should at least demonstrate that they have the intended
behavior for all testcases.

\subsubsection{Transformation Performance}

Secondly, runtime performance should be evaluated (using the performance
testcases). An online spreadsheet will be 
provided for conveniently entering and comparing performance results. Solution
submitters are also encouraged to reflect on the good (or poor) performance
of their solution, compared to solutions from TTC'13 competitors.

\subsubsection{Transformation Understandability}

Peer reviews will be used to assess qualitatively the understandability
of all solutions. We envision online reviews involving multiple rounds
in order to reach consensus among all participants.

\subsubsection{Reproducibility}

Solution submitters should facilitate the online review by (1) providing
an appendix that lists all code, diagrams, scripts, pop-ups that are
essential for running the transformation, (2) publishing all artifacts
to SHARE. Point (1) ensures that reviewers know where to look while
(2) ensures that reviewers can put all artifacts in the proper perspective.
For (2), solution submitters are encouraged to provide on the VM desktop
one shortcut per testcase. Ideally, clicking such a shortcut automatically
demonstrates that the test is passed by the solution. However, this
is hard since we do not offer any infrastructure for automatically
analyzing whether an output EMF file is the same as the testcase-provided
one. Solution submitters who have technology for automating the test
procedure are encouraged to share that via the TTC'13 forum. As case
proposers, we at least offer the aforementioned GMF editors with an
automatic layouting feature for the effective human-based analysis
of the generated output models.

\subsection{Bonus Criteria}

\subsubsection{Verification Support}

Solution submitters who have technologies to formalize the class of
Petri-Nets supported by the reduction rules are encouraged to demonstrate
how such technologies should be used effectively. Eshuis has already
defined the class of supported nets mathematically~\cite{eshuis09fm},
but a computer-supported, formal link between the definition of that
class and the definition of the reduction rules is missing. Once the
definition of the class of supported input models is linked formally
to the transformation rules of a solution, it becomes interesting
to reason about properties that are satisfied by the set of rules.
For example, it would be interesting to have automated analysis features
for verifying whether or not the rules terminate for the complete
class defined by Eshuis~\cite{eshuis09fm}. Similarly, an automated
confluence analysis of the rules set would be highly valuable too.
It is important that such analyses are applied to classes of input
models rather than just to individual inputs. Once automated support
for verifying the termination and confluence of a rule set is in place,
one can more efficiently experiment with variations to the reduction
rules. Finally, note that it would also be useful to reverse the direction
of the formal analysis: if it turns out that given a formally defined
class of input models $c$, properties $P$ (e.g., termination and
confluence) are \emph{not} satisfied for a set of rules $R$, it would
be useful to learn for which subclasses $S$ of $c$ the properties
$P$ would be satisfied nonetheless. With the latter type of support
in place, it would be interesting to compare the ``reverse engineered''
formal definition of a maximal subclass against the ``manually formalized''
class of supported nets that is described in~\cite{eshuis09fm}.

\subsubsection{Simulation Support}

Solution submitters are also encouraged to demonstrate features related
to the execution of the Petri-Nets and their corresponding statecharts.
Note however that execution of the models in isolation is not considered
relevant to this case. Instead, we are interested in tools that facilitate
for example the simulation of statecharts based on a rule-based Petri-Net
simulator (or vice versa).

\subsubsection{Transformation Language support for Change Propagation}

We are interested in typical scenario's of change propagation (automatically
creating/updating/deleting output model elements when performing updates
on a previously transformed input model). To the best of our knowledge,
most graph and transformation languages do not yet offer built-in
support for this so experts are encouraged to show their unique features
for this use case. To facilitate comparison between solutions, we encourage 
solution developers to address the following change propagation scenarios.
However, there may be additional change propagation scenarios that are well-suited 
to a particular tool and solution developers may also describe these scenarios
in addition or instead of the suggestions below.

\begin{itemize}
  \item Adding a place and a transition in the input Petri-Net should result
        in a corresponding state and hyper-edge in the output statechart. The
        arcs should also be updated accordingly. All other elements should remain
        unchanged. No new AND or OR states should be created.
  \item Updating the name of a place or transition that was previously transformed
        should only result in the update of the corresponding statechart element. 
        Furthermore, removing a place (transition) followed by adding a new place
        (transition) should not be misinterpreted as a renaming operation when 
        propagating changes.
  \item Removing a place and transition from a sequence in the input Petri-Net should
        result in the removal of the corresponding state and hyper-edge in the output
        statechart. The arcs should also be updated accordingly. All other elements 
        should remain unchanged.
\end{itemize}

Test cases can be provided on request for change propagation scenarios (including
but not limited to those listed above).

\subsubsection{Transformation Language support for Reversing the Transformation}

Previous editions of the TTC did not yet attract convincing solutions
based on languages with support for bidirectional transformations.
We consider PN2SC a particularly challenging case for such languages
since the reduction rules are input-destructive. Therefore, we especially
encourage experts in such languages to check whether they can solve
the case and also report potentially any negative results.

\subsubsection{Transformation Tool support for Debugging}

For validation, specialized debuggers tend to offer features for visualizing
(input, intermediate and output) models, ideally in the syntax of
the corresponding input or output language. Breakpoints can typically
be set at the rule level but more fine-grained control can be provided
too. Solution submitters are encouraged to document at which level
their transformation tool supports breakpoint definition. Also, they
are encouraged to document any features they may have employed for
turning a generic model/graph/object debugger into a Petri-Net and/or
statechart-specific debugger.

\subsubsection{Transformation Tool support for Refactoring}

Since transformation definitions have to evolve, automated restructuring
(refactoring) operations are very relevant but very little transformation
tools provide them at the time of writing. This may hinder their adoption
compared to general purpose programming languages. Therefore, solution
submitters are also encouraged to document any support their tool
suite offers for the systematic restructuring of transformation definitions.

\section{Table for Comparing Solutions}

In order to facilitate the comparison of all submitted results, we
encourage solution submitters to document their solution in terms
of Table~\ref{tab:Table-for-Results}. A solution name should distinguish
a solution from competing solutions that are based on the same languages
or tools (if any). The column ``Performance Optimizations'' should
clarify techniques that were employed for optimizing performance:
\emph{N} indicates that no optimizations were used, \emph{M} indicates
that an optimization algorithm was implemented by hand and\emph{ E}
indicates that declarative (engine-provided) optimizations were employed.
The table also provides columns for entering the results of the performance
tests, as measured within a SHARE VM. The SHARE URL should be provided
as a bibtex citation to minimize the size of the table. The column
``Understandability'' will be based upon the results of the peer reviews,
in a qualitative scale\emph{ (very poor, poor, neutral, good, excellent)}.
Finally the table includes columns for concisely describing the results
for bonus criteria. For 5.2.1, \emph{C} stands for some kind of automatic
\emph{confluence} analysis, \emph{T} stands for some kind of automatic
\emph{termination} analysis while \emph{N} stands for \emph{no} automatic
verification support. The specific type of verification support should
be elaborated in free text.

\begin{table}
\begin{sideways}
\begin{tabular}{|c|c|c|c|c|c|c|c|c|c|c|c|c|c|c|c|c|c|c|c|}
\hline 
{\tiny Solution } & {\tiny Language} & {\tiny Language } & {\tiny Language } & {\tiny Performance } & {\tiny sp200} & {\tiny sp1000} & {\tiny sp5000} & {\tiny sp10000} & {\tiny sp40000} & {\tiny sp200000} & {\tiny SHARE URL} & {\tiny Understandability} & \begin{sideways}
{\tiny Bonus 5.2.1}
\end{sideways} & \begin{sideways}
{\tiny Bonus 5.2.2}
\end{sideways} & \begin{sideways}
{\tiny Bonus 5.2.2}
\end{sideways} & \begin{sideways}
{\tiny Bonus 5.2.3}
\end{sideways} & \begin{sideways}
{\tiny Bonus 5.2.4}
\end{sideways} & \begin{sideways}
{\tiny Bonus 5.2.5}
\end{sideways} & \begin{sideways}
{\tiny Bonus 5.2.6}
\end{sideways}\tabularnewline
{\tiny Name} & {\tiny{} for Initialization } & {\tiny for Reduction } & {\tiny for Orchestration} & {\tiny Optimizations} &  &  &  &  &  &  &  &  &  &  &  &  &  &  & \tabularnewline
\hline 
\hline 
 &  &  &  & {\tiny N / M / E } & {\tiny ... ms} & {\tiny ... ms} & {\tiny ... ms} & {\tiny ... ms} & {\tiny ... ms} & {\tiny ... ms} &  & {\tiny -2 .. +1} & {\tiny C/T/N } & {\tiny Y/N} & {\tiny Y/N} & {\tiny Y/N} & {\tiny Y/N} & {\tiny Y/N} & {\tiny Y/N}\tabularnewline
 &  &  &  &  &  &  &  &  &  &  &  &  &  &  &  &  &  &  & \tabularnewline
\hline 
\end{tabular}
\end{sideways}

\caption{\label{tab:Table-for-Results}Table for concisely documenting a solution.}

\end{table}

\vspace{5mm}
 \textbf{Acknowledgements}: The authors thank Prof. Juan de Lara for
constructing the initial versions of the Petri-Net and statecharts
metamodels used in this case.

\bibliography{pn2sc}

\begin{thebibliography}{10}
\providecommand{\bibitemdeclare}[2]{}
\providecommand{\surnamestart}{}
\providecommand{\surnameend}{}
\providecommand{\urlprefix}{Available at }
\providecommand{\url}[1]{\texttt{#1}}
\providecommand{\href}[2]{\texttt{#2}}
\providecommand{\urlalt}[2]{\href{#1}{#2}}
\providecommand{\doi}[1]{doi:\urlalt{http://dx.doi.org/#1}{#1}}
\providecommand{\bibinfo}[2]{#2}

\bibitemdeclare{incollection}{vAmstel2011ATLQVTperfAnal}
\bibitem{vAmstel2011ATLQVTperfAnal}
\bibinfo{author}{Marcel \surnamestart van Amstel\surnameend},
  \bibinfo{author}{Steven \surnamestart Bosems\surnameend},
  \bibinfo{author}{Ivan \surnamestart Kurtev\surnameend} \&
  \bibinfo{author}{Luís \surnamestart Ferreira~Pires\surnameend}
  (\bibinfo{year}{2011}): \emph{\bibinfo{title}{Performance in Model
  Transformations: Experiments with {ATL} and {QVT}}}.
\newblock In \bibinfo{editor}{Jordi \surnamestart Cabot\surnameend} \&
  \bibinfo{editor}{Eelco \surnamestart Visser\surnameend}, editors: {\sl
  \bibinfo{booktitle}{Theory and Practice of Model Transformations}}, {\sl
  \bibinfo{series}{Lecture Notes in Computer Science}} \bibinfo{volume}{6707},
  \bibinfo{publisher}{Springer Berlin / Heidelberg}, pp.
  \bibinfo{pages}{198--212}, \doi{10.1007/978-3-642-21732-6\_14}.

\bibitemdeclare{article}{Bacon1994compilertrans}
\bibitem{Bacon1994compilertrans}
\bibinfo{author}{David~F. \surnamestart Bacon\surnameend},
  \bibinfo{author}{Susan~L. \surnamestart Graham\surnameend} \&
  \bibinfo{author}{Oliver~J. \surnamestart Sharp\surnameend}
  (\bibinfo{year}{1994}): \emph{\bibinfo{title}{Compiler transformations for
  high-performance computing}}.
\newblock {\sl \bibinfo{journal}{ACM Comput. Surv.}}
  \bibinfo{volume}{26}(\bibinfo{number}{4}), pp. \bibinfo{pages}{345--420},
  \doi{10.1145/197405.197406}.

\bibitemdeclare{techreport}{eshuis05pn2scTechreport}
\bibitem{eshuis05pn2scTechreport}
\bibinfo{author}{Rik \surnamestart Eshuis\surnameend} (\bibinfo{year}{2005}):
  \emph{\bibinfo{title}{{S}tatecharting {P}etri {N}ets}}.
\newblock \bibinfo{type}{Technical Report} \bibinfo{number}{Beta WP 153},
  \bibinfo{institution}{Eindhoven University of Technology}.
\newblock \urlprefix\url{http://beta.ieis.tue.nl/node/1267}.

\bibitemdeclare{inproceedings}{eshuis09fm}
\bibitem{eshuis09fm}
\bibinfo{author}{Rik \surnamestart Eshuis\surnameend} (\bibinfo{year}{2009}):
  \emph{\bibinfo{title}{Translating Safe Petri Nets to Statecharts in a
  Structure-Preserving Way}}.
\newblock In \bibinfo{editor}{Ana \surnamestart Cavalcanti\surnameend} \&
  \bibinfo{editor}{Dennis \surnamestart Dams\surnameend}, editors: {\sl
  \bibinfo{booktitle}{{FM}}}, {\sl \bibinfo{series}{{LNCS}}}
  \bibinfo{volume}{5850}, \bibinfo{publisher}{Springer}, pp.
  \bibinfo{pages}{239--255}, \doi{10.1007/978-3-642-05089-3\_16}.
\newblock \bibinfo{note}{Extended as Beta WP 282 at Eindhoven University of
  Technology}.

\bibitemdeclare{inproceedings}{EshuisVanGorp2012ER}
\bibitem{EshuisVanGorp2012ER}
\bibinfo{author}{Rik \surnamestart Eshuis\surnameend} \&
  \bibinfo{author}{Pieter \surnamestart {Van Gorp}\surnameend}
  (\bibinfo{year}{2012}): \emph{\bibinfo{title}{Synthesizing Object Life Cycles
  from Business Process Models}}.
\newblock In \bibinfo{editor}{Paolo \surnamestart Atzeni\surnameend},
  \bibinfo{editor}{David~W. \surnamestart Cheung\surnameend} \&
  \bibinfo{editor}{Sudha \surnamestart Ram\surnameend}, editors: {\sl
  \bibinfo{booktitle}{{ER}}}, {\sl \bibinfo{series}{Lecture Notes in Computer
  Science}} \bibinfo{volume}{7532}, \bibinfo{publisher}{Springer}, pp.
  \bibinfo{pages}{307--320}, \doi{10.1007/978-3-642-34002-4\_24}.

\bibitemdeclare{inproceedings}{Feather87progtranssurv}
\bibitem{Feather87progtranssurv}
\bibinfo{author}{M.~S. \surnamestart Feather\surnameend}
  (\bibinfo{year}{1987}): \emph{\bibinfo{title}{A survey and classification of
  some program transformation approaches and techniques}}.
\newblock In: {\sl \bibinfo{booktitle}{The IFIP TC2/WG 2.1 Working Conference
  on Program specification and transformation}},
  \bibinfo{publisher}{North-Holland Publishing Co.},
  \bibinfo{address}{Amsterdam, The Netherlands, The Netherlands}, pp.
  \bibinfo{pages}{165--195}.

\bibitemdeclare{proceedings}{TTC11Proceedings}
\bibitem{TTC11Proceedings}
\bibinfo{editor}{Pieter~Van \surnamestart Gorp\surnameend},
  \bibinfo{editor}{Steffen \surnamestart Mazanek\surnameend} \&
  \bibinfo{editor}{Louis \surnamestart Rose\surnameend}, editors
  (\bibinfo{year}{2011}): \emph{\bibinfo{title}{Proceedings Fifth
  Transformation Tool Contest}}. {\sl
  \bibinfo{series}{EPTCS}}~\bibinfo{volume}{74}, \doi{10.4204/EPTCS.74}.

\bibitemdeclare{inproceedings}{Gronmo2009_ComparisonofThreeModelTransformation%
Languages}
\bibitem{Gronmo2009_ComparisonofThreeModelTransformationLanguages}
\bibinfo{author}{R.~\surnamestart Gr{\o}nmo\surnameend},
  \bibinfo{author}{B.~\surnamestart M{\o}ller-Pedersen\surnameend} \&
  \bibinfo{author}{G.K. \surnamestart Olsen\surnameend} (\bibinfo{year}{2009}):
  \emph{\bibinfo{title}{Comparison of Three Model Transformation Languages}}.
\newblock In: {\sl \bibinfo{booktitle}{ECMDA-FA'09: European Conference on
  Model Driven Architecture - Foundations and Applications}}, {\sl
  \bibinfo{series}{LNCS}} \bibinfo{volume}{5562},
  \bibinfo{publisher}{Springer}, pp. \bibinfo{pages}{2--17},
  \doi{10.1007/978-3-642-02674-4\_2}.

\bibitemdeclare{incollection}{Lohmann09BPMtransSurv}
\bibitem{Lohmann09BPMtransSurv}
\bibinfo{author}{Niels \surnamestart Lohmann\surnameend}, \bibinfo{author}{Eric
  \surnamestart Verbeek\surnameend} \& \bibinfo{author}{Remco \surnamestart
  Dijkman\surnameend} (\bibinfo{year}{2009}): \emph{\bibinfo{title}{Petri Net
  Transformations for Business Processes --- A Survey}}.
\newblock In \bibinfo{editor}{Kurt \surnamestart Jensen\surnameend} \&
  \bibinfo{editor}{Wil~M. \surnamestart Aalst\surnameend}, editors: {\sl
  \bibinfo{booktitle}{Transactions on Petri Nets and Other Models of
  Concurrency {II}}}, \bibinfo{publisher}{Springer-Verlag},
  \bibinfo{address}{Berlin, Heidelberg}, pp. \bibinfo{pages}{46--63},
  \doi{10.1007/978-3-642-00899-3\_3}.

\bibitemdeclare{article}{Rensink2010_Graphtransformationtoolcontest2008}
\bibitem{Rensink2010_Graphtransformationtoolcontest2008}
\bibinfo{author}{Arend \surnamestart Rensink\surnameend} \&
  \bibinfo{author}{Pieter \surnamestart Van~Gorp\surnameend}
  (\bibinfo{year}{2010}): \emph{\bibinfo{title}{Graph transformation tool
  contest 2008}}.
\newblock {\sl \bibinfo{journal}{International Journal on Software Tools for
  Technology Transfer (STTT)}} \bibinfo{volume}{12}, pp.
  \bibinfo{pages}{171--181}, \doi{10.1007/s10009-010-0157-7}.

\bibitemdeclare{inproceedings}{rose10comparison}
\bibitem{rose10comparison}
\bibinfo{author}{L.M. \surnamestart Rose\surnameend},
  \bibinfo{author}{M.~\surnamestart Herrmannsdoerfer\surnameend},
  \bibinfo{author}{J.R. \surnamestart Williams\surnameend},
  \bibinfo{author}{D.S. \surnamestart Kolovos\surnameend},
  \bibinfo{author}{K.~\surnamestart Garc\'{e}s\surnameend},
  \bibinfo{author}{R.F. \surnamestart Paige\surnameend} \&
  \bibinfo{author}{F.A.C. \surnamestart Polack\surnameend}
  (\bibinfo{year}{2010}): \emph{\bibinfo{title}{A Comparison of Model Migration
  Tools}}.
\newblock In \bibinfo{editor}{D.C. \surnamestart Petriu\surnameend},
  \bibinfo{editor}{N.~\surnamestart Rouquette\surnameend} \&
  \bibinfo{editor}{{\O}~\surnamestart Haugen\surnameend}, editors: {\sl
  \bibinfo{booktitle}{{MODELS}'10: International Conference on Model Driven
  Engineering Languages and Systems}}, {\sl \bibinfo{series}{LNCS}}
  \bibinfo{volume}{6394}, \bibinfo{publisher}{Springer}, pp.
  \bibinfo{pages}{61--75}, \doi{10.1007/978-3-642-16145-2\_5}.

\bibitemdeclare{inproceedings}{Rose2009_AnAnalysisofApproachestoModelMigration}
\bibitem{Rose2009_AnAnalysisofApproachestoModelMigration}
\bibinfo{author}{Louis~M. \surnamestart Rose\surnameend},
  \bibinfo{author}{Richard~F. \surnamestart Paige\surnameend},
  \bibinfo{author}{Dimitrios~S. \surnamestart Kolovos\surnameend} \&
  \bibinfo{author}{Fiona~A.C. \surnamestart Polack\surnameend}
  (\bibinfo{year}{2009}): \emph{\bibinfo{title}{An Analysis of Approaches to
  Model Migration}}.
\newblock In: {\sl \bibinfo{booktitle}{{MoDSE-MCCM}'09: Joint {MoDSE-MCCM}
  Workshop on Models and Evolution}}, pp. \bibinfo{pages}{6--15}.
\newblock
  \urlprefix\url{http://www.modse.fr/modsemccm09/doku.php?id=Proceedings}.

\bibitemdeclare{inproceedings}{Taentzer2005_ModelTransformationsbyGraphTransfo%
rmationsAComparativeStudy}
\bibitem{Taentzer2005_ModelTransformationsbyGraphTransformationsAComparativeSt%
udy}
\bibinfo{author}{Gabriele \surnamestart Taentzer\surnameend},
  \bibinfo{author}{Karsten \surnamestart Ehrig\surnameend},
  \bibinfo{author}{Esther \surnamestart Guerra\surnameend},
  \bibinfo{author}{Juan~De \surnamestart Lara\surnameend},
  \bibinfo{author}{Tihamer \surnamestart Levendovszky\surnameend},
  \bibinfo{author}{Ulrike \surnamestart Prange\surnameend} \&
  \bibinfo{author}{Daniel \surnamestart Varro\surnameend}
  (\bibinfo{year}{2005}): \emph{\bibinfo{title}{Model Transformations by Graph
  Transformations: A Comparative Study}}.
\newblock In: {\sl \bibinfo{booktitle}{Model Transformations in Practice
  Workshop at MODELS 2005, Montego}}, p.~\bibinfo{pages}{05}.

\bibitemdeclare{article}{Tolosa2011TSMatlMeasurement}
\bibitem{Tolosa2011TSMatlMeasurement}
\bibinfo{author}{Jos\'e~Barranquero \surnamestart Tolosa\surnameend},
  \bibinfo{author}{Oscar \surnamestart Sanju\'an-Mart\'inez\surnameend},
  \bibinfo{author}{Vicente \surnamestart Garc\'ia-D\'iaz\surnameend},
  \bibinfo{author}{B.~Cristina~Pelayo \surnamestart G-Bustelo\surnameend} \&
  \bibinfo{author}{Juan Manuel~Cueva \surnamestart Lovelle\surnameend}
  (\bibinfo{year}{2011}): \emph{\bibinfo{title}{Towards the systematic
  measurement of {ATL} transformation models}}.
\newblock {\sl \bibinfo{journal}{Softw. Pract. Exper.}} \bibinfo{volume}{41},
  pp. \bibinfo{pages}{789--815}, \doi{10.1002/spe.1033}.

\bibitemdeclare{misc}{vangorp10PNtoHSCiiii}
\bibitem{vangorp10PNtoHSCiiii}
\bibinfo{author}{Pieter \surnamestart {Van Gorp}\surnameend}
  (\bibinfo{year}{2013}): \emph{\bibinfo{title}{Online demo: {PN2SC} in {J}ava
  and {G}r{G}en}}.
\newblock
  \bibinfo{howpublished}{\url{http://share20.eu/?page=ConfigureNewSession\&vdi%
=XP\_GB9\_GrGen\_live\_AD2HSC\_i\_i\_i\_i\_i\_i\_Epsilon\_GrGenTTC13\_i.vdi}}.

\bibitemdeclare{inproceedings}{VanGorp2010MoDELS}
\bibitem{VanGorp2010MoDELS}
\bibinfo{author}{Pieter \surnamestart {Van Gorp}\surnameend} \&
  \bibinfo{author}{Rik \surnamestart Eshuis\surnameend} (\bibinfo{year}{2010}):
  \emph{\bibinfo{title}{Transforming process models: executable rewrite rules
  versus a formalized java program}}.
\newblock In: {\sl \bibinfo{booktitle}{Proceedings of the 13th international
  conference on Model driven engineering languages and systems: Part II}},
  \bibinfo{series}{MODELS'10}, \bibinfo{publisher}{Springer-Verlag},
  \bibinfo{address}{Berlin, Heidelberg}, pp. \bibinfo{pages}{258--272},
  \doi{10.1007/978-3-642-16129-2\_19}.

\bibitemdeclare{inproceedings}{Varro2008_TransformationofUMLModelstoCSPACaseSt%
udyforGraphTransformationTools}
\bibitem{Varro2008_TransformationofUMLModelstoCSPACaseStudyforGraphTransformat%
ionTools}
\bibinfo{author}{D\'aniel \surnamestart Varr\'o\surnameend},
  \bibinfo{author}{M\'ark \surnamestart Asztalos\surnameend},
  \bibinfo{author}{D\'enes \surnamestart Bisztray\surnameend},
  \bibinfo{author}{Artur \surnamestart Boronat\surnameend},
  \bibinfo{author}{Duc-Hanh \surnamestart Dang\surnameend},
  \bibinfo{author}{Rubino \surnamestart Gei{\ss}\surnameend},
  \bibinfo{author}{Joel \surnamestart Greenyer\surnameend},
  \bibinfo{author}{Pieter \surnamestart Van~Gorp\surnameend},
  \bibinfo{author}{Ole \surnamestart Kniemeyer\surnameend},
  \bibinfo{author}{Anantha \surnamestart Narayanan\surnameend},
  \bibinfo{author}{Edgars \surnamestart Rencis\surnameend} \&
  \bibinfo{author}{Erhard \surnamestart Weinell\surnameend}
  (\bibinfo{year}{2008}): \emph{\bibinfo{title}{Transformation of {UML} Models
  to {CSP}: A Case Study for Graph Transformation Tools}}.
\newblock In: {\sl \bibinfo{booktitle}{{AGTiVE}'08: International Symposium on
  Applications of Graph Transformation with Industrial Relevance}}, {\sl
  \bibinfo{series}{{LNCS}}} \bibinfo{volume}{5088},
  \bibinfo{publisher}{Springer}, pp. \bibinfo{pages}{540--565},
  \doi{10.1007/978-3-540-89020-1\_36}.

\end{thebibliography}
 \bibliographystyle{eptcs}
\end{document}